\documentclass[]{article}
\usepackage{amsmath}
\usepackage{amsfonts}
\usepackage{amssymb}
\usepackage{graphicx}
\usepackage{amsfonts}
\usepackage[english]{babel}
\usepackage[utf8]{inputenc}
\usepackage{times}
\usepackage[T1]{fontenc}

\begin{document}

\title{Uncertainty Principle in Loop Quantum Cosmology by Moyal Formalism}

\author{Leonid Perlov, \\
Department of Physics, University of Massachusetts,  Boston, USA\\
leonid.perlov@umb.edu
}

\maketitle

\begin{abstract}
In this paper we derive the uncertainty principle for the Loop Quantum Cosmology homogeneous and isotropic FLWR model with the holonomy-flux algebra. The uncertainty principle is between the variables $c$, with the meaning of connection and $\mu$ having the meaning of the physical cell volume to the power $2/3$, i.e $v^{2/3}$  or a plaquette area. Since both $\mu$ and $c$ are not operators, but rather the random variables, the Robertson uncertainty principle derivation that works for hermitian operators, can not be used. Instead we use the  Wigner-Moyal-Groenewold phase space formalism. The  Wigner-Moyal-Groenewold formalism was originally applied to the Heisenberg algebra of the Quantum Mechanics. One can derive from it both the canonical and path integral QM as well as the uncertainty principle. In this paper we apply it to the holonomy-flux algebra in case of the homogeneous and isotropic space. Another result  is the expression for the  Wigner function on the space of the cylindrical wave functions defined on $R_b$ in $c$ variables rather than in dual space $\mu$ variables. 
\end{abstract}
 
\section{Introduction}
We derive the uncertainty principle for the loop quantum cosmology variables $c$ having the meaning of the connection and $\mu$ having the meaning of the physical cell volume to the power $2/3$, i.e $v^{2/3}$ $\cite{Lewandowski}$ p.242 or a plaquette area. Since both $c$ and $\mu$ are not operators but rather the random variables used in the holonomy operator, we can not use the general Heisenberg inequality straightforward approach, which works for the hermitian operators. Instead we use the Wigner-Moyal-Groenewold  $\cite{Moyal}$ $\cite{Wigner}$ $\cite{Groenewold}$ phase space approach. The Moyal's statistical approach can be used to derive the quantum mechanics both in the canonical and path integral forms. It also provides the way to obtain the uncertainty principle by using the cumulants and cumulant generating function. The Wigner-Moyal-Groenewold approach can be applied to derive the uncertainty principle for the Loop Quantum Cosmology if one uses the holonomy-flux algebra instead of the Heisenberg algebra, i.e. $[\hat{p}, \hat{N}] = a \mu \hat{N}$ instead of $[\hat{p}, \hat{q}] = - i\hbar$, where 
a - constant,  $ a = \frac{8 \pi \gamma G \hbar}{3}$, $\hat{N}$ is an LQC holonomy operator $\cite{AshtekarLQC}, \cite{Bojowald} $ $\hat{N}_{(\mu)} = e^{i\mu c}$, where $c$ - is the configuration variable corresponding to the connection, $\mu$ - the number of the fiducial cell repetition, also it has the second meaning of physical cell volume to the power $2/3$, i.e  $v^{2/3}$ $\cite{Lewandowski}$ p.242 or a plaquette area, $ \mu \in R, \; c \in R_b$ - Bohr real line compactification. 
With the assumption that Immirzi parameter $\gamma$ is real we obtain:
\begin{equation}
\sigma^2_c \sigma^2_{\mu} \ge \left (\frac{4 \pi \gamma G}{3} \right )^2 \frac{\hbar^2}{4}
\end{equation}

The paper is organized as follows. In section $\ref{sec:CharacteristicFunction}$ we derive the holonomy-flux characteristic function. In section $\ref{sec:UncertaintyPrinciple}$ we obtain the LQC uncertainty principle. The discussion section $\ref{sec:Discussion}$ concludes the paper. In the Appendix A we present the theorems with proofs for the LQC Wigner function and its momentum properties. 

\section{Holonomy-Flux Characteristic Function}
\label{sec:CharacteristicFunction}
In order to derive the LQC uncertainty principle for the random variables $c$ and $\mu$ by using the Wigner-Moyal-Groenewold formalism, we need to derive first the LQC Wigner function. The LQC Wigner function in the dual space variables $\mu$ was previously obtained in $\cite{Fewster}$, however no derivation for the best of our knowledge was completed for the LQC Wigner function in the original variables  $c$.  We are going to derive it in such form and prove in this section its two important properties, which will be used for the uncertainty principle derivation in the next section. The first property - when integrated by one variable the Wigner function becomes the distribution density of the other variable.  The second is that all momentum integrals used in Moyal derivation still exists when integration is performed with respect to the holonomy-flux measures.\\[2ex] 
The holonomy-flux algebra in case of the homogeneous isotropic space FLWR model can be described by using the commutation relation between the holonomy operator and the momentum operator:  \cite{AshtekarLQC}\cite{Bojowald}\cite{BojowaldBook}
\begin{equation}
[\hat{N}_{(\mu)}, \hat{p}] = -\frac{8 \pi \gamma G \hbar}{3}\mu \hat{N}_{(\mu)}
\end{equation}
The holonomy and flux operators act as follows: \\
\begin{equation}
\label{HolonomyFluxOperators}
\hat{N}_{(\mu)}\Psi(c) = e^{i\mu c}\Psi(c),  \quad  \hat{p}\Psi(c) = -i \frac{8\pi \gamma G \hbar}{3}\frac{d\Psi}{dc}
\end{equation}
The basis of the physical Hilbert space is given by LQC analogs of LQG spin-networks: $\hat{N}_{(\mu)} = e^{i\mu c}$, where $c$ - is the configuration variable corresponding to the connection, $\mu$ - the number of the fiducial cell repetition also having the meaning of physical cell volume to the power $2/3$, i.e $v^{2/3}$ $\cite{Lewandowski}$ or a plaquette area, $c \in R_b$ - Bohr compactified real line, $\mu \in R$.\\
The basis functions satisfy the relation:
\begin{equation}
\langle N_{(\mu)}|N_{\mu'}\rangle =  \langle e^{i\mu c} e^{i\mu' c} \rangle = \delta_{\mu , \mu'}
\end{equation}
The FLWR  holonomy-flux algebra commutator is of the form:
\begin{equation}
\label{HolonomyFlux}
[\hat p, \hat N] = a \mu \hat N
\end{equation}
,where $a$ is a constant:
\begin{equation}
\label{a}
a = \frac{4 \pi \gamma G \hbar}{3}
\end{equation}
We are going to obtain the LQC characteristic function $M(\tau, \theta)$ and its inverse Fourier transform.\\
Let us formally define the LQC Winger function as:
\begin{equation}
\label{Fmuc1}
F(\mu, c) = \int {\psi}^{*}(c - a \tau)e^{-2 ia \tau \mu} \psi(c + a \tau) d\tau
\end{equation}
where $\psi(c)$ are the cylindrical functions of the $c \in R_b$, compactified real line. The cylindrical functions are of the form:\\
\begin{equation}
\label{Psi}
\psi(c) = \sum\limits_{n=0}^{N} {\hat{\Psi}_{\mu_n}e^{i\mu_n c}}, \;\; {\mu}_n \in R
\end{equation}
In order for $F(\mu, c)$ to have the meaning of the mutual quasi distribution function of $\mu$ and $c$ the following two equalities should be true. When integrated with respect to one variable it becomes the distribution density of the other one: 
\begin{equation}
\label{FmucInt1}
\rho_c = \int F(\mu, c) d\mu = |\psi(c)|^2
\end{equation}
and
\begin{equation}
\label{FmucInt2}
\rho_{\mu} = \int F(\mu, c) dc = |\hat{\Psi}_{\mu}|^2
\end{equation}
In order to prove both equalities we use measures $dc$ and $d\mu$ as in \cite{Fewster} \cite{Rudin} \cite{Thiemann}.
\begin{equation}
\int\limits_{\hat{R_b}} \hat{f_{\mu}} d\mu = \sum\limits_{\mu \in R} \hat{f_{\mu}}
\end{equation}
\begin{equation}
\label{MeasureC1}
\int\limits_{R_b}  e^{i\mu c} \; dc = \delta_{\mu, 0}
\end{equation}
,where $\hat{R_b}$ - is Bohr's dual space, $ \delta_{\mu, 0}$ - a Kronecker delta.\\
The characters of the compactified line $R_b$ are the functions $h_{\mu}(c) = e^{i\mu c}$ \cite{Rudin}.
The Fourier transform of the function on $R_b$ is given by : 
\begin{equation}
\label{MeasureC}
\hat{f_{\mu}} = \int f(c) h_{-\mu}(c) dc
\end{equation}
This is an isomorphism of $L^2(R_b, c) \rightarrow L^2(\hat{R_b}, d\mu)$. $e^{i\mu c}$ comprise the basis of $H = L^2(R_b, dc)$.
The proof of the equalities ($\ref{FmucInt1}$) and ($\ref{FmucInt2}$) is given in the Theorems 1 and 2 in the Appendix A. These properties imply that $F(\mu, c)$ is an LQC Wigner function in $(c, \mu)$ variables. \\
In order to use for the derivation of the LQC uncertainty principle we would need to prove one more equality - the expression for the first momentum: 
\begin{equation}
\label{Fmuc3}
\int F(\mu, c)  e^{2 ia \tau_0 \mu} d \mu =  \psi^*(c - a\tau_0) \psi(c + a \tau_0) 
\end{equation}
,where $ c, \tau_0 \in R_b $\\
or by replacing the variables one can also write it in the form:
\begin{equation}
\label{Fmuc3ab}
\int F(\mu, c)  e^{ ia \tau_0 \mu} d\mu =
\psi^*(c - \frac{a\tau_0}{2}) \psi(c + \frac{a \tau_0}{2})
\end{equation}
The proof of this equality is provided in the Theorem 3 of the Appendix A.

\section{Loop Quantum Cosmology Uncertainty Principle}
\label{sec:UncertaintyPrinciple}
We now have all necessary tools ready to derive the LQC uncertainty principle for $c$ and $\mu$ random variables in the manner similar to Moyal's derivation for the Heisenberg algebra $\cite{Moyal}$.\\
The distribution  $\rho(c)$ as obtained in ($\ref{FmucInt1}$) is
\begin{equation}
\label{rho}
\rho(c) = \int F(\mu, c) \; d\mu = \psi(c) \psi^*(c)
\end{equation}
We define the characteristic function $M(\tau|c)$ of $\tau$ conditional in c.
\begin{equation}
\label{LQCChar}
M( \tau | c) = \frac{1}{\rho} \int F(\mu, c) e^{i\tau \mu} d\mu
\end{equation}
By substituting ($\ref{Fmuc3ab}$) and ($\ref{rho}$) into ($\ref{LQCChar}$) we obtain:
\begin{equation}
M( \tau | c) = \frac{1}{\rho} \int F(\mu, c) e^{i\tau \mu} d\mu =  \frac{ \psi^*(c - \frac{a \tau}{2} )\;\psi(c + \frac{a \tau}{2})}{\psi^*(c) \psi(c)}
\end{equation}
Following Moyal formalism $\cite{Moyal}$ we replace the variables with the new ones - the amplitude and the phase:
\begin{equation}
\psi(c) = \rho(c)^{\frac{1}{2}} e^{iS(c)/\hbar}
\end{equation}
The cumulant function $\cite{Kendall}$ of $M( \tau | c )$ is:
\begin{multline}
\label{cumulantfunction}
K(\tau | c ) = \log M(\tau, c)= \frac{1}{2} \log \rho( c + \frac{a \tau}{2} ) + \frac{1}{2}  \log \rho(c - \frac{a \tau}{2} ) + \\
  -\log(\rho(c)) 
+ \frac{i}{\hbar}\left [S( c + \frac{a \tau}{2} ) - S( c - \frac{a \tau}{2} )\right]
\end{multline}
The cumulants (coefficients of $\frac{{(i\tau)}^n}{n!}$ in the Taylor expansion of $K(\tau | c )$) are:
\begin{equation}
\bar{k}_{1}(c) = \frac{1}{i}\frac{\partial K(\tau | c)}{\partial \tau}|_{\tau = 0} = \frac{a}{2 \hbar}  \frac{\partial S(c)}{\partial c}  -  \frac{-a}{2 \hbar} \frac{\partial S(c)}{\partial c} = \frac{a}{\hbar}  \frac{\partial S(c)}{\partial c}
\end{equation}
\begin{equation}
\label{K2}
 \bar{k}_{2}(c) =  \sigma^2_{p|c}  = \frac{1}{i^2}\frac{\partial^2 K(\tau | c)}{\partial^2 \tau}|_{\tau = 0}= - \frac{a^2}{4} {\left ( \frac{\partial ^2 \log \rho(c)}{\partial c^2} \right ) }  
\end{equation}
The meaning of the first cumulant for the normal distribution is the expected value, while the second is the variance $\cite{Kendall}$:
\begin{equation}
\bar{k}_1(c) = \overline{\overline{\mu}}, \quad  \bar{k}_2(c) = \sigma^2_{\mu|c} = \overline{ \overline{\mu^2}} - (\overline{\overline{\mu}})^2 
\end{equation}
The derivation is similar to Moyal $\cite{Moyal}$, however instead of Heisenberg algebra cumulants we use the holonomy-flux algebra cumulants $\bar{k}_{1}(c)$ and $\bar{k}_2(c)$ to obtain the LQC uncertainty principle.\\
For the two random variables $\alpha$ and  $\beta$ with zero means we write the Cauchy-Schwarz-Bunyakovsky inequality:
\begin{equation}
\label{Cauchy}
|(\overline{\alpha^2}\; \overline{\beta^2}| = \sigma_{\alpha} \sigma_{\beta} \ge |\overline{\alpha\beta}|
\end{equation}
Taking $\alpha = \overline{\overline{\mu}}$ and $\beta = c$ and assuming $\overline{\mu} = \overline{c} = 0$, we obtain from ($\ref{Cauchy}$) :
\begin{equation}
\label{Cauchy2}
\sigma_c \sigma(\overline{\overline{\mu}}) \ge \left | \int c \overline{\overline{\mu}} \; \rho(c) \; dc \right | = |\overline{c \;\mu} |
\end{equation}
Now taking:
\begin{equation}
\label{Label1}
\alpha = \frac{\partial \log \rho }{\partial c}, \quad \bar{\alpha} = \int \frac{\partial \log \rho }{\partial c} \; \rho \; dc = 0
\end{equation}
we can write
\begin{equation}
\label{alpha2}
\overline{\alpha^2} = \int \left ( \frac{\partial \log \rho }{\partial c} \right )^2 \; \rho \; dc = -\int \frac{\partial^2 \log \rho}{\partial c^2} \; \rho \; dc 
\end{equation}
By expressing $\frac{\partial^2 \log \rho}{\partial c^2} $ through $ \sigma^2_{\mu|c}$ from ($\ref{K2}$) we obtain:
\begin{equation}
-\frac{\partial^2 \log \rho}{\partial c^2}  = \frac{4}{a^2}  \sigma^2_{\mu|c}  
\end{equation}
Then by substituting it into ($\ref{alpha2}$) we get:
\begin{equation}
\label{alphasquare}
\overline{\alpha^2} = -\int \frac{\partial^2 \log \rho}{\partial c^2} \; \rho \; dc
= \int \frac{4}{a^2} \sigma^2_{\mu|c}  \; \rho \; dc 
\end{equation}
and 
\begin{equation}
\overline{\alpha c} = \int c \frac{\partial \log \rho }{\partial c} \; \rho \; dc = -1
\end{equation}
by multiplying ($\ref{alphasquare}$) by $\sigma^2_c$ and using the Cauchy-Schwarz-Bunyakovsky inequality ($\ref{Cauchy}$) where $\beta = c$ and using
\begin{equation}
| \overline{\sigma^2_c} \; \overline{\alpha^2} | \ge |\overline{\alpha c}| = 1
\end{equation}
we obtain:
\begin{equation}
\overline{\sigma^2_c} \; \overline{\alpha^2} = \frac{4 \sigma^2_c }{a^2} \int  \sigma^2_{\mu|c}  \; \rho \; dc \ge 1
\end{equation}
by assuming that Immirzi $\gamma$ is real and therefore $a^2 > 0 $, we get
\begin{equation}
\label{label3}
\sigma^2_c \int   \sigma^2_{\mu|c}  \rho \; dc  \ge \frac{a^2}{4}
\end{equation}
By noticing that 
\begin{equation}
\label{sigmap}
\sigma^2_{\mu} = \int \left ( \sigma^2_{\mu|c} + (\overline{\overline{\mu}})^2 \right ) \; \rho \; dc
\end{equation}
by multiplying ($\ref{sigmap}$) by $\sigma^2_c$ from the left we obtain:
\begin{multline}
\sigma^2_c \sigma^2_{\mu} = \sigma_c^c \int  \sigma_{\mu|c}^2 \; \rho \; dc + \sigma_c^2  \int  (\overline{\overline{\mu}})^2 \; \rho \; dc  =  \\
 \sigma_c^2 \int  \sigma_{\mu|c}^2 \; \rho \; dc + \sigma_c^2 \sigma^2(\overline{\overline{\mu}}) \; \ge  \sigma_c^2  \int \sigma_{\mu|c}^2 \; \rho \; dc + (\overline{c\mu})^2
\end{multline}
,where in the last inequality above we used ($\ref{Cauchy2}$).
By dropping the last term we can rewrite the inequality as:
\begin{equation}
\sigma^2_c\sigma^2_{\mu} \ge \sigma^2_c  \int \sigma^2_{\mu|c} \; \rho \; dc
\end{equation}
Finally by using ($\ref{label3}$) we obtain:
\begin{equation}
\sigma^2_c \sigma^2_{\mu} \ge \frac{a^2}{4}  
\end{equation}
By remembering that expression for $ a = \frac{4 \pi \gamma G \hbar}{3}$
we can rewrite it as:
\begin{equation}
\sigma^2_c \sigma^2_{\mu} \ge \left (\frac{4 \pi \gamma G}{3} \right )^2 \frac{\hbar^2}{4}
\end{equation}

\section{ Discussion }
\label{sec:Discussion}
By using Moyal's approach for the holonomy-flux algebra, we have obtained the uncertainty principle for the Loop Quantum Cosmology in the case of homogeneous and isotropic space: 
\begin{equation}
\sigma^2_c \sigma^2_{\mu} \ge \left (\frac{4 \pi \gamma G}{3} \right )^2 \frac{\hbar^2}{4}
\end{equation}
Even though the variables $c$ and $\mu$ are the random variables and not the operators, the uncertainty principle still exists. It can not be derived by the Robertson uncertainty principle approach, which is applicable only to the hermitian operators, however as we have shown it can be done by using the statistical Moyal-Wigner-Groenewold approach for the random variables. Another novel result of this paper is the expression for the  Wigner function (\ref{Fmuc1})  on the space of the cylindrical wave functions defined on $R_b$ in $c$ with the demonstrated properties. 
\\[4ex]
\textbf{Acknowledgment}\\
I would like to specially thank Michael Bukatin for the multiple fruitful and challenging discussions. \\[2ex]
This paper is dedicated to the memory of Evgeny Moiseev, who originally introduced me to the Einstein's theory of relativity.  \\[3ex]

\section{Appendix A \quad LQC Wigner Function Properties }
\label{sec:AppendixA}
\textbf{Theorem 1:}\\[2ex]
\begin{equation}
\label{FmucInt1a}
\rho_c = \int F(\mu, c) d\mu = |\psi(c)|^2
\end{equation}\\[2ex]
\textbf{Proof}:\\[2ex]
 We substitute the expression ($\ref{Fmuc1}$) of $F(\mu, c)$ and  the expression ($\ref{Psi}$) for $\psi(c)$ into ($\ref{FmucInt1a}$).
\begin{equation}
\label{FmucInt3}
\int F(\mu, c) d\mu = \int \int \sum\limits_{n=0}^{N} \sum\limits_{k=0}^{K}  {\hat{\Psi}^*_{\mu_n}e^{-ia \mu_n c}}e^{ia \mu_n \tau} {\hat{\Psi}_{\mu_k}e^{ia \mu_k c}}e^{ia \mu_k \tau} e^{-2 ia \tau \mu}  d \tau \; d \mu
\end{equation}
,where $\tau \in R_b, \; \mu \in R$. The integration with respect to $\mu$ is just a sum as $\mu$ is discrete. 
\begin{equation}
\label{FmucInt5}
\int F(\mu, c) d\mu = \sum_{\mu \in R} \sum\limits_{n=0}^{N} \sum\limits_{k=0}^{K}  \int {\hat{\Psi}^*_{\mu_n}e^{-ia \mu_n c}}e^{ia \mu_n \tau} {\hat{\Psi}_{\mu_k}e^{ia \mu_k c}}e^{ia \mu_k \tau} e^{-2 ia \tau \mu}  d \tau 
\end{equation}
By collecting the terms containing $\tau$ and integrating with respect to $\tau$, by using ($\ref{MeasureC1}$) we obtain:
\begin{equation}
\int e^{ia \mu_n \tau} e^{ia \mu_k \tau} e^{-2 ia \tau \mu}  d \tau = \delta_{2 \mu, \mu_k + \mu_n}
\end{equation}
Since $\mu \in R$, summation by $\mu$ makes the terms with  $2 \mu \ne \mu_k + \mu_n$ equal zero and the terms with  $\mu = \mu_k + \mu_n$ equal one and all terms with $\tau$  and $\mu$ disappear from the sum. In other words for each pair $\mu_n$ and $\mu_k$ there exists $\mu$ such that  $2 \mu = \mu_k + \mu_n$  and that $\mu$ keeps the $\mu_k$ and $\mu_n$ in the sum, all other terms with $\tau$ zero out in the integration and we obtain:
\begin{equation}
\label{FmucInt11}
\int F(\mu, c) d\mu =  \sum\limits_{n=0}^{N} \sum\limits_{k=0}^{K}  {\hat{\Psi}^*_{\mu_n}e^{-ia \mu_n c}} {\hat{\Psi}_{\mu_k}e^{ia \mu_k c}} = \psi^*(c) \psi(c) = |\psi(c)|^2
\end{equation}
${\square}$\\[4ex]
\textbf{Theorem 2:}\\[2ex]
\begin{equation}
\label{FmucInt2a}
\rho_{\mu} = \int F(\mu, c) dc = |\hat{\Psi}_{\mu}|^2
\end{equation}\\[2ex]
\textbf{Proof:}\\[2ex]
In order to prove this equality we substitute the expression ($\ref{Fmuc1}$) of $F(\mu, c)$ and  the expression ($\ref{Psi}$) for $\psi(c)$ into ($\ref{FmucInt2a}$).
\begin{equation}
\label{FmucInt4}
\int F(\mu, c)  dc = \int \int \sum\limits_{n=0}^{N} \sum\limits_{k=0}^{K}  {\hat{\Psi}^*_{\mu_n}e^{-ia \mu_n c}}e^{ia \mu_n \tau} {\hat{\Psi}_{\mu_k}e^{ia \mu_k c}}e^{ia \mu_k \tau} e^{-2 ia \tau \mu}  d \tau \; dc
\end{equation}
,where $c, \tau \in R_b, \; \mu \in R$\\
The integration with measure $dc$ by ($\ref{MeasureC1}$) gives:
\begin{equation}
\label{mu1}
\int e^{-ia \mu_n c} e^{ia \mu_k c} dc =  \delta_{\mu_k - \mu_n, 0} 
\end{equation}
Therefore only the terms with $\mu_k = \mu_n$ remain in the sums in ($\ref{FmucInt4}$). The integration with respect to $d \tau$ in turn gives
\begin{equation}
\label{mu5}
\int e^{ia \mu_n \tau} e^{ia \mu_k \tau} e^{-2 ia \tau \mu} d\tau = \delta_{2\mu, \mu_k + \mu_n}
\end{equation}
From ($\ref{mu1}$) and ($\ref{mu5}$) it follows that:
\begin{equation}
\mu_n = \mu_k = \mu
\end{equation}
after substituting it into ($\ref{FmucInt4}$) we obtain:
\begin{equation}
\label{FmucInt5}
\int F(\mu, c)  dc = \int \int  {\hat{\Psi}^*_{\mu}e^{-ia \mu c}}e^{ia \mu \tau} {\hat{\Psi}_{\mu}e^{ia \mu c}}e^{ia \mu \tau} e^{-2 ia \tau \mu}  d \tau \; dc
\end{equation}
the integrals with respect to $d\tau$ and $dc$ are equal to one according to ($\ref{MeasureC1}$), so ($\ref{FmucInt5}$) becomes:
\begin{equation}
\label{Psimu}
\int F(\mu, c)  dc =  {\hat{\Psi}^*_{\mu}} {\hat{\Psi}_{\mu}}  = |\hat{\Psi}_{\mu}|^2
\end{equation}
${\square}$\\[4ex]
\textbf{Theorem 3:}\\[2ex]
\begin{equation}
\label{Fmuc3a}
\int F(\mu, c)  e^{2 ia \tau_0 \mu} d \mu =  \psi^*(c - a\tau_0) \psi(c + a \tau_0) 
\end{equation}
,where $ c, \tau_0 \in R_b $\\[2ex]
\textbf{Proof:}\\[2ex]
We begin by substituting the expression of $F(\mu, c)$  ($\ref{Fmuc1}$) into the l.h.s. of ($\ref{Fmuc3a}$)
\begin{equation}
\label{Fmuc4}
\int F(\mu, c)  e^{2 ia \tau_0 \mu} d \mu = \int \int {\psi}^{*}(c - a \tau)e^{-2 ia \tau \mu} e^{2 ia \tau_0 \mu}\psi(c + a \tau) d\tau \; d\mu
\end{equation}
By using the expression ($\ref{Psi}$) for the $\psi(c)$ function, we obtain:
\begin{equation}
\label{FmucInt6}
\int F(\mu, c)  e^{2 ia \tau_0 \mu} d\mu = \int \int \sum\limits_{n=0}^{N} \sum\limits_{k=0}^{K}  {\hat{\Psi}^*_{\mu_n}e^{-ia \mu_n c}}e^{ia \mu_n \tau} {\hat{\Psi}_{\mu_k}e^{ia \mu_k c}}e^{ia \mu_k \tau} e^{-2 ia \tau \mu}  e^{2 ia \tau_0 \mu} d \tau \; d \mu
\end{equation}
again, the integration by $\mu$ can be replaced with the sum over $\mu$
\begin{equation}
\label{FmucInt7}
\int F(\mu, c)  e^{2 ia \tau_0 \mu} d\mu = \sum\limits_{\mu \in R} \sum\limits_{n=0}^{N} \sum\limits_{k=0}^{K} \int {\hat{\Psi}^*_{\mu_n}e^{-ia \mu_n c}}e^{ia \mu_n \tau} {\hat{\Psi}_{\mu_k}e^{ia \mu_k c}}e^{ia \mu_k \tau} e^{-2 ia \tau \mu}  e^{2 ia \tau_0 \mu} d \tau 
\end{equation}
The integration by $\tau$ gives us as before:
\begin{equation}
\label{mu2}
\int e^{ia \mu_n \tau} e^{ia \mu_k \tau} e^{-2 ia \tau \mu} d\tau = \delta_{2\mu, \mu_k + \mu_n}
\end{equation}
Which means that only those $\mu$ satisfying $2\mu = \mu_k + \mu_n$  are equal to one after the integration, all the rest are zeros and we obtain:
\begin{equation}
\label{FmucInt8}
\int F(\mu, c)  e^{2 ia \tau_0 \mu} d\mu =  \sum\limits_n \sum\limits_k  {\hat{\Psi}^*_{\mu_n}e^{-ia \mu_n c}} {\hat{\Psi}_{\mu_k}e^{ia \mu_k c}}   e^{2 ia \tau_0 \mu} 
\end{equation}
We substitute $2 \mu = \mu_k + \mu_n$ into ($\ref{FmucInt8}$) and use the definition of the LQC cylindrical functions ($\ref{Psi}$):
\begin{equation}
\label{FmucInt9}
\int F(\mu, c)  e^{2 ia \tau_0 \mu} d\mu =  \sum\limits_n \sum\limits_k  {\hat{\Psi}^*_{\mu_n}e^{-ia \mu_n c}} {\hat{\Psi}_{\mu_k}e^{ia \mu_k c}}   e^{ ia \tau_0 (\mu_n + \mu_k) }  = \psi^*(c - a\tau_0) \psi(c + a \tau_0)
\end{equation}
or by taking $\tau_0/2$ instead of $\tau_0$ it can be rewritten in the form:
\begin{equation}
\label{FmucInt10}
\int F(\mu, c)  e^{ ia \tau_0 \mu} d\mu =  \sum\limits_n \sum\limits_k  {\hat{\Psi}^*_{\mu_n}e^{-ia \mu_n c}} {\hat{\Psi}_{\mu_k}e^{ia \mu_k c}}   e^{ \frac{i a \tau_0 (\mu_n + \mu_k)}{2} } =
\psi^*(c - \frac{a\tau_0}{2}) \psi(c + \frac{a \tau_0}{2})
\end{equation}
This completes the proof of the equality ($\ref{Fmuc3a}$)\\
${\square}$\\[4ex]

\end{document}